\documentclass[12pt,preprint]{aastex}

\newcommand{\pks}{PKS~2155--304}
\newcommand{\xmm}{{\it XMM-Newton}}
\newcommand{\sax}{{\it BeppoSAX}}

\newcommand{\gr}{$\gamma$-ray}

\newcommand{\et}{et al.\ }

\shortauthors{Zhang}
\shorttitle{Signature of IC X-ray Emission in PKS~2155--304}

\begin{document}

\title{XMM-Newton observations of the TeV BL Lac object PKS~2155--304 in 2006: signature of inverse Compton X-ray emission?}

\author{Y.H. Zhang}
\affil{Department of Physics and Tsinghua Center for Astrophysics (THCA), Tsinghua University, Beijing 100084, China}
\email{youhong.zhang@mail.tsinghua.edu.cn}

                       %-------------------------%
\begin{abstract}

This paper reports the first discovery of possible inverse Compoton (IC) X-ray emission below 10~keV in the typical high-energy peaked BL Lac object (HBL) PKS~2155--304. Two {\it XMM-Newton} observations performed in 2006 reveal that the 0.6--10~keV X-ray spectra of the source harden ($\Delta\Gamma \sim$~0.1--0.3) at break energies of $\sim4$~keV. The concave X-ray spectra of the source could be easily interpretated by a mixture of a steep component (i.e., the high energy tail of the synchrotron emission) and a flat one (i.e., the low energy side of the IC emission). However, the steep spectra ($\Gamma \sim$~2.3) in the hard X-rays indicate that the synchrotron emission still domiantes over the IC one, while the latter is effectively present and perceived as flattening the synchrotron spectrum in this energy range. The quasi-simultaneous optical-UV-X-ray spectral energy distributions (SEDs) obtained with \xmm\ suggest that the concave X-ray spectra of the source could be the result of down-ward shift of the synchrotron peak frequency to the optical band, incuring the IC emisssion to become more important in the hard ($\sim4$--10~keV) X-ray band with respective to other cases in which the synchrotron emission peaks in the UV-soft-X-ray range. This discovery provides a new clue for a smooth transition between HBLs and the low-energy peaked BL Lac objects (LBLs).

\end{abstract}


                %--------------------%

\section{Introduction}\label{sec:intro}

BL Lac objects are currently classified on the basis of the peak frequency of synchrotron emission. The synchrotron emission of the high-energy peaked BL Lac objects (HBLs) peaks in the UV-X-ray range, while the low-energy peaked BL Lac objects (LBLs) are those whose synchrotron emission peaks in the IR-optical band. The transition from HBLs to LBLs is expected to be smooth rather than dichotomic, forming a continuous sequence of synchrotron peak frequency with luminosity. A few bright LBLs, when they are in very high states, can exhibit X-ray spectra typical of HBLs.

\pks\ ($z$=0.116) is the brightest BL Lac object at the UV and EUV wavelengths as revealed by the IUE and EUVE satellites. It is also among the brightest X-ray BL Lac objects. Though it is classified as a HBL, the peak energy of its synchrotron emission were found to be no larger than a few tenth of a keV even in very high states (Zhang \et 2002; Massaro et al. 2008). Numerous X-ray observations with various X-ray telescopes instead suggested that the synchrotron emission of \pks\ tends to peak in the UV-EUV rather than in the X-rays. In contrast, for the other two well studied HBLs Mrk~421 and Mrk~501, the synchrotron emission can peak at a few~keV or even at very high energy X-ray band (e.g., Pian \et 1998; Massaro \et 2004; Tramacere et al. 2007). It is the fifth among the twenty extragalactic very high energy (VHE) or TeV \gr\ sources known so far (Chadwick \et 1999). An extreme TeV outburst of about an hour duration was recorded on 2006 July 28, and the most rapid TeV variation was seen down to a timescale of $\sim 200$~s (Aharonian \et 2007). However, the rapid TeV flare did not accompany spectral variability, and SWIFT follow-up observations did not reveal shift of the synchrotron peak frequency either (Foschini et al. 2007). The CANGAROO-III observations also showed raipd VHE \gr\ variability over timescales shorter than a few hours (Sakamoto et al. 2008).   

As a calibration target, \xmm\ repeatedly monitored \pks\ since 2000. In this paper, we present X-ray spectral analysis and quasi-simultaneous optical-UV-X-ray SEDs of two observations performed on 2006 May 1 (orbit 1171, obs. ID 0158961401) and November 7 (orbit 1266, obs. ID 0411780101), to show the signature of the possible IC emission between $\sim$~4 and 10~keV. A complete data analysis for all available \xmm\ obervations will be presented elsewhere.

                   %---------------------------------%

\section{The XMM-Newton Observations and Data Reduction}\label{sec:obs}

We follow the standard procedures to reprocess the Observation Data Files (ODF) with the \xmm\ Science Analysis System (SAS) version 7.1.0 and with the calibration files as of 2007 October. Firstly we extract a high energy (10~keV~$< E >$~12~keV for pn, and $E >$~10~keV for mos) light curve only with single event (PATTERN=0) from full frame of the exposed CCD, to identify intervals of flaring particle background for pn, mos1 and mos2, respectively. No background flares are found. Pile-up effects are then examined with the SAS task {\it epatplot}. The pn data are not affected by the pile-up effects, while the mos data meet with strong pile-up effects. As a result, we extract pn source counts from a circle region centered on the source with radius of 40 arcsec, while the mos1/mos2 source counts are extracted from an annulus region centered on the source with inner and outer radii of 5/10 and 40 arcsec, respectively. The pn background counts are selected from a circle region least affected by the source counts on the same CCD as the source region. Because there are no mos background region available on the same CCD as the source region, we select mos background counts from a circle region of another exposed CCD. Single and double (pn) and single-to-quadruple (mos) events with quality FLAG=0 are selected for our analysis. For consistency, we use the same energy range of 0.6--10~keV for both pn and mos data, in an attempt to exclude any possible remaining (cross-)calibration uncertainties below 0.6~keV. The pn and mos spectra are created with SAS task XMMSELECT and grouped with FTOOL task GRPPHA in order to have at least 30 counts in each energy bin for the use of $\chi^2$ statistics. Redistribution matrices and ancillary response files are produced with SAS task RMFGEN and ARFGEN. The journal of the pn and mos observations is shown in Table~\ref{tab:epic}. 

The Optical Monitor (OM) observations were taken in standard imaging mode with a sequence of six filters (i.e., V,B,U, UVW1, UVM2 and UVW2). The OM observations have two images for UVW1 filter and one image for other filters. We reprocess the OM imaging data using the standard \textit{omichain} pipeline of SAS 7.1.0, from which we obtain the source count rates and instrumental magnitudes. The OM count rates are then converted into fluxes according to the average conversion factors derived from observations of white dwarf standard stars. The OM observational results are listed in Table~\ref{tab:om}.  

Part of the \xmm\ observations obtained on 2002 November 29-30 (orbit 0545, obs. ID 0124930601) for \pks\ are also reprocessed in the same way as above, which is mainly used to compare with the observations in 2006. Table~\ref{tab:epic} and \ref{tab:om} also show the observational journal of the orbit 0545.

                     %-----------------------------%

\section{X-ray data analysis}\label{sec:xray}

\subsection{Timing analysis}\label{sec:lc}

\begin{figure}
\plottwo{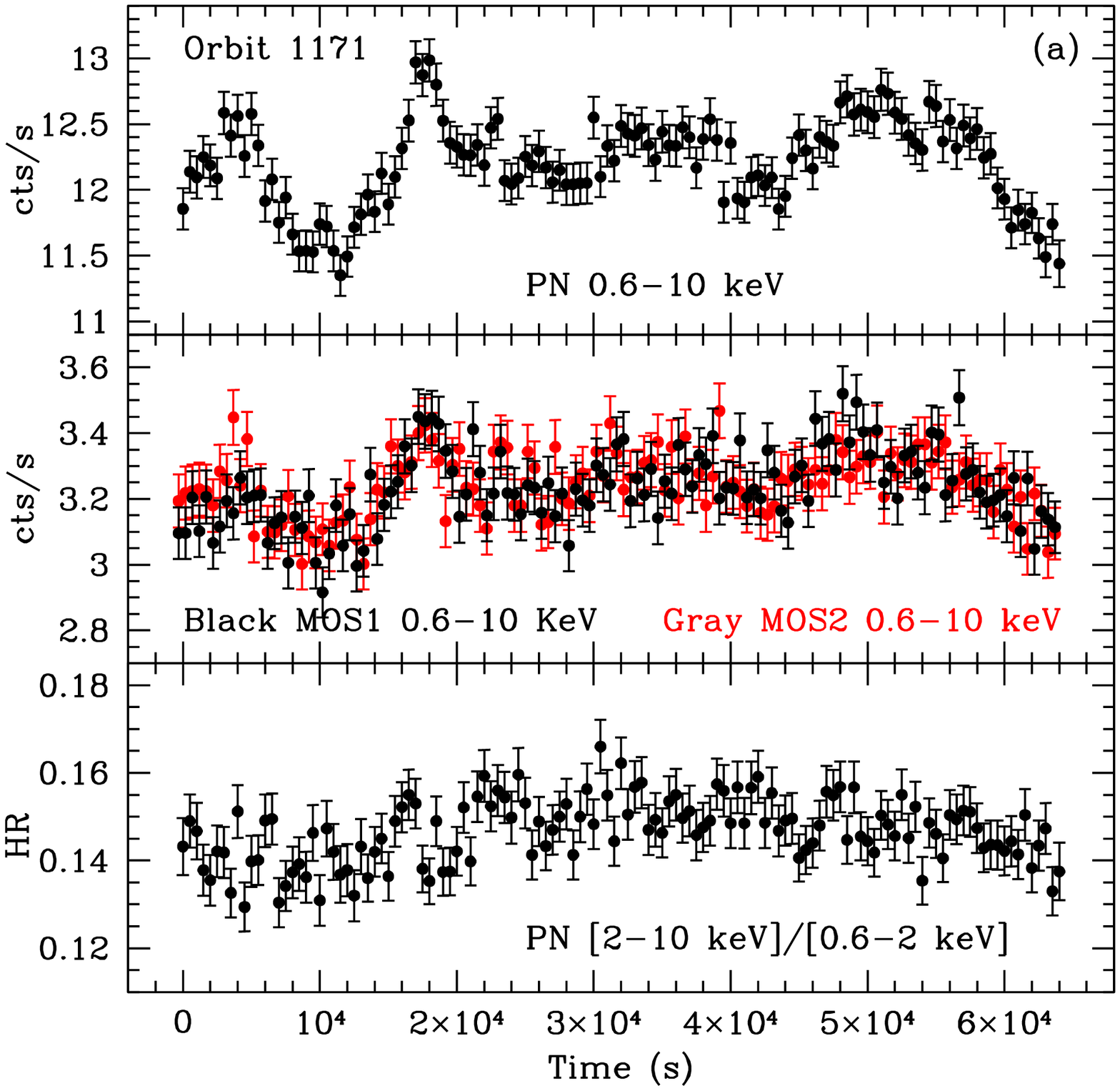}{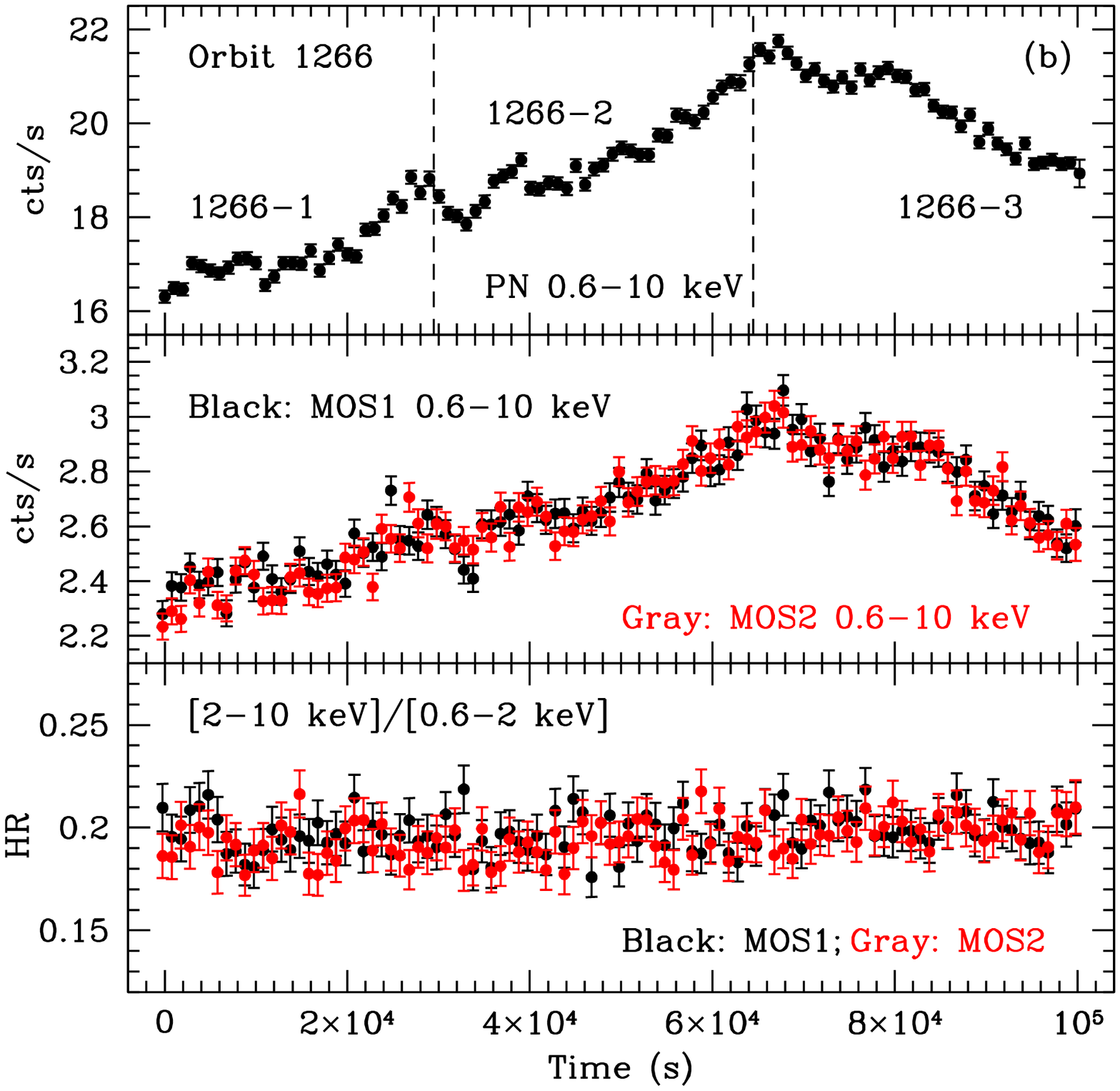}
\caption { \footnotesize  
(a) Orbit 1171. The 500s binned 0.6--10~keV pn (top pannel) and mos (middle pannel) light curves, and the pn hardness ratios of 2--10 to 0.6--2~keV (bottom pannel). Both pn and mos were operated in small window with medium filter.
(b) Orbit 1266. The 1000s binned 0.6--10~keV pn (top pannel) and mos (middle pannel) light curves, and the mos hardness ratios of 2--10 to 0.6--2~keV (bottom pannel). Pn were operated in small window with different filters, seperating the pn light curve into three sections (distinguished by the dashed lines), i.e., 1266-1 (thin), 1266-2 (medium) and 1266-3 (thick), and resulting offsets between them, particularly between 1266-2 and 1266-3. Sections 1266-2 and 1266-3 have been arbitrarily scaled up in order to have a visual view of a ``complete'' pn light curve. Mos were operated in small window with medium filter. The pn and mos light curves appear to track each other after arbitrarily scaling the three sections of the pn light curve together.
} 
\label{fig:lc}
\end{figure}

Figure~\ref{fig:lc}a shows the 0.6--10~keV light curves and hardness ratios for orbit 1171. The pn light curve exhibits an overall variation of $\sim 10\%$ from the minimum to the maximum, characterized by multiple ``flare-like'' events. The mos1 and mos2 light curves are basically identical, and both roughly tracks the pn one. The pn hardness ratios (2--10 versus 0.6--2~keV) are not strongly variable.   

The 0.6--10~keV light curves and hardness ratios for orbit 1266 are plotted in Figure~\ref{fig:lc}b. The pn light curve is separated into three sections due to different filters used for 1266-1 (thin), 1266-2 (medium) and 1266-3 (thick), respectively. This brings about an offset between different sections, in particular between 1266-2 and 1266-3. The mos1 and mos2 light curves overlap each other very well, showing an asymetric flare with $\sim 30\%$ minimum-to-maximum variability amplitude and a rising timescale of $\sim$~70~ks. The decaying phase of the flare may be not fully sampled, but its variating rate is slower than the rising one. The mos1 and mos2 hardness ratios (2--10 to 0.6--2~keV) are roughly constant. The pn light curve appears to follow the mos ones very well after one arbitrarily scales the three sections of the pn light curve together (see Figure~\ref{fig:lc}b).

\subsection{Spectral analysis}\label{sec:spectra}

Roughly constant hardness ratios imply that the X-ray spectra are not strongly variable during each orbit, so we fit a (mean) spectrum with XSPEC 12.3.1 for each exposure. The Galactic neutral hydrogen absorption ($N_H = 1.36 \times 10^{20}$~${\rm cm^{-2}}$, Lockman \& Savage 1995) in sight of light toward \pks\ is fixed for all of spectral fits. The results of the spectral fits are summarized in Table~\ref{tab:fit}.

\begin{figure}
\centering
\includegraphics[scale=0.31, angle=270]{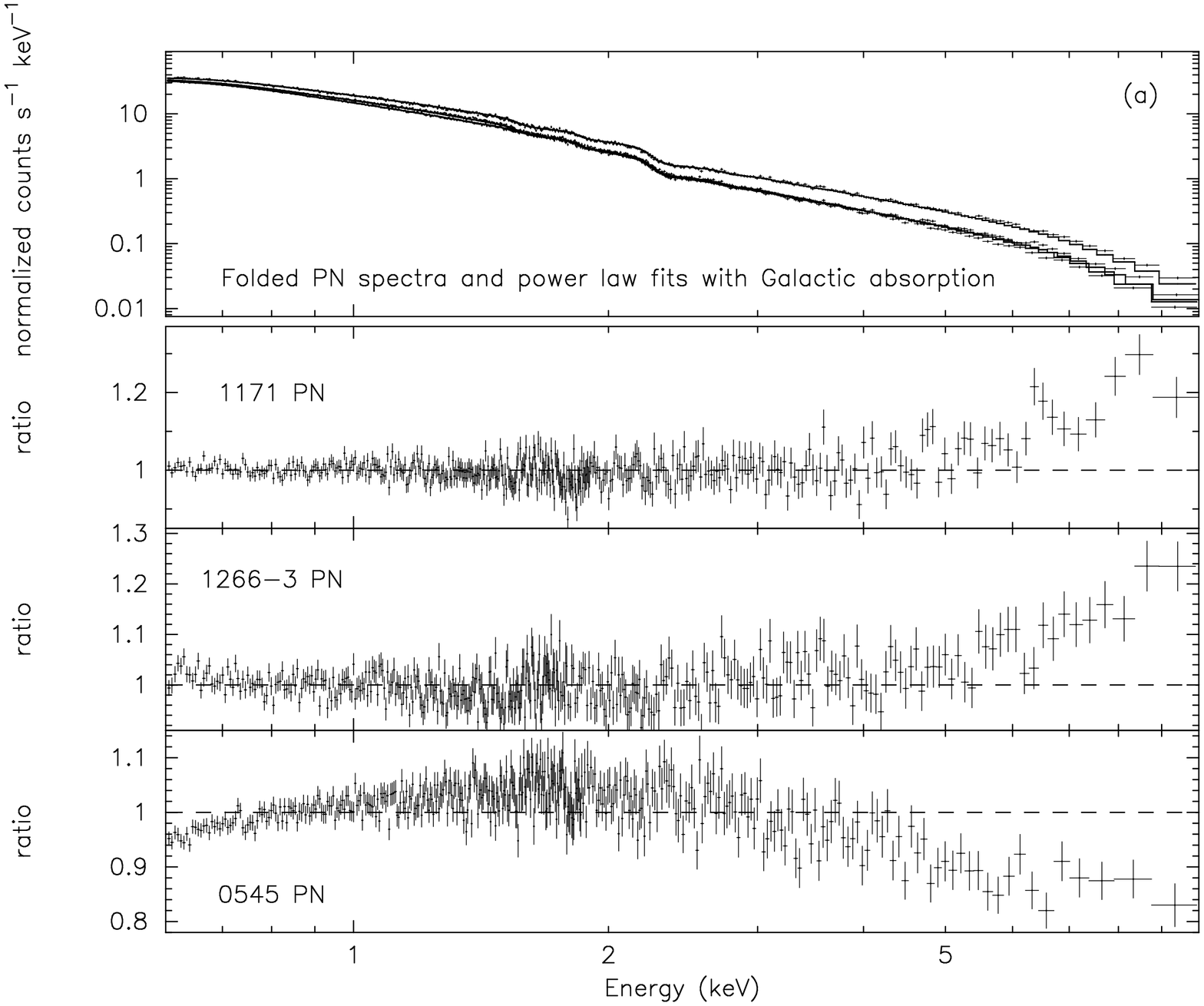}\includegraphics[scale=0.31, angle=270]{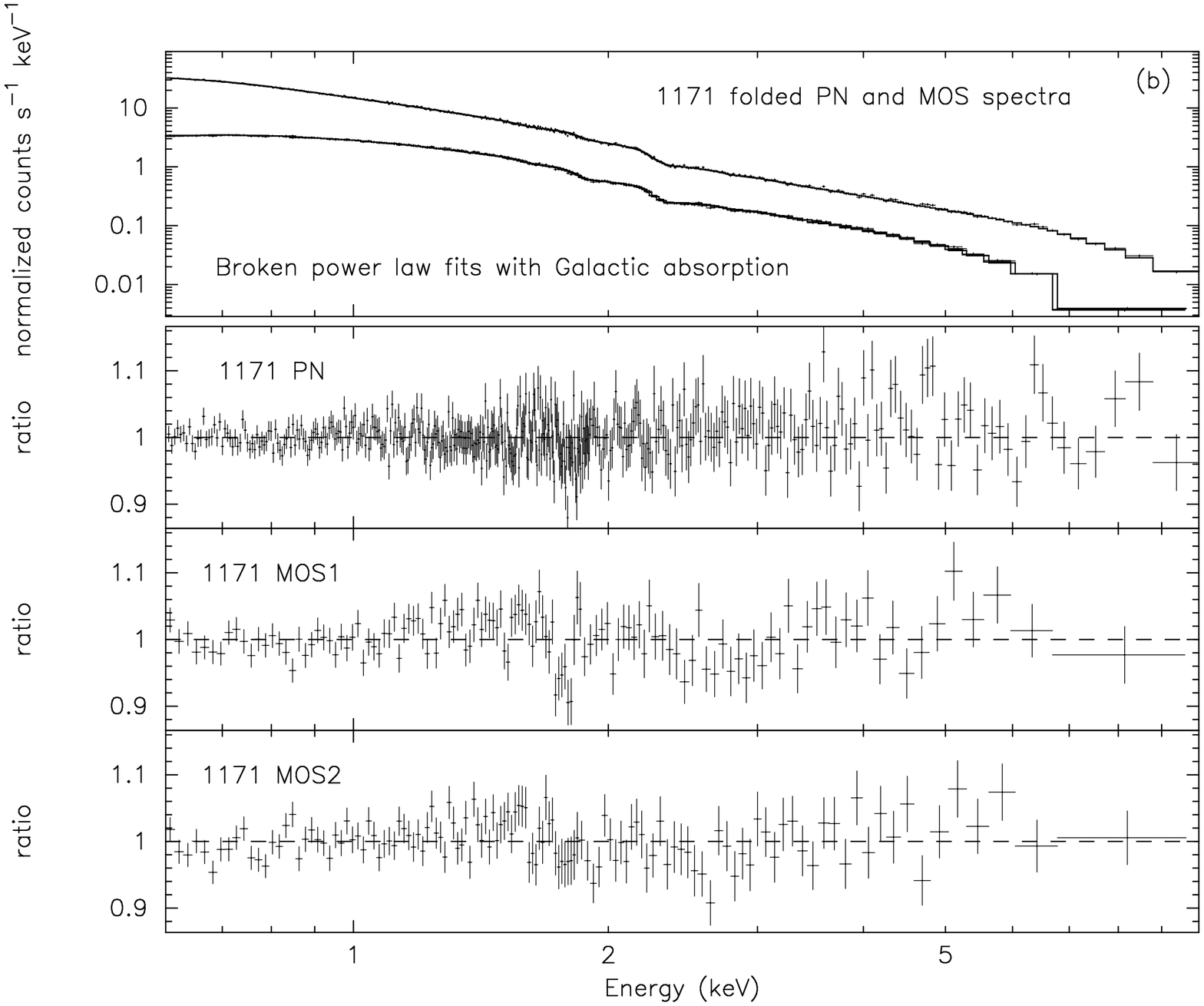}
\caption {\footnotesize  
(a) The pn count spectra of orbit 1171, 1266-3 and 0545, and respective best power-law fit model (top panel). The data-to-model ratios (2-4 panels) clearly show the upturn of the spectra for orbit 1171 and 1266-3, and continuously down-ward curved spectrum for orbit 0545. 
(b) The pn, mos1 and mos2 count spectra for orbit 1171 with respective best broken power-law fit model (top panel) and the data-to-model ratios (2-4 panels).  
} 
\label{fig:ratio}
\end{figure}

Firstly we fit each pn spectrum with a single power-law model to inspect if the spectrum is typical of HBLs, characterized by continuously downward-curved (convex) shape with increasing energy. The pn count spectra with the best power-law fit models and the data-to-model ratios are shown in Figure~\ref{fig:ratio}a for orbit 1171 and 1266-3, respectively. To our surprise, both fits clearly show positive residuals above $\sim$~4--5~keV, becoming larger with increasing energy. This feature indicates a spectral flattening with increasing energy. A single power-law fit leaves similar residuals for the other two pn spectra (i.e., the orbit 1266-1 and 1266-2) and all the mos spectra. Due to much smaller photon counts compared to the pn ones, the residuals for the mos spectra are relatively weak.

It is well known that the residuals, from a single power-law fit to the X-ray spectra of HBLs, are negative below a few tenth of keV and above a few keV, and are positive between them, implying that the spectra are continuously down-ward curved below 10~keV. The broad band X-ray spectra with \sax\ suggest that a spectral hardening of \pks\ occurs above tens of keV, while the spectra below 10~keV are still typical of HBLs (Giommi et al. 1998; Chiappetti et al. 1999). To firmly believe the reality of our new discovery, we make several tests. A single power-law fit to the pn spectra extracted only with single events (minimizing the pile-up effects) yields identical photon index and residuals. We then extract the pn spectra from an annulus by excluding the inner 10 arcsec to avoid probable residuals of the pile-up effects, and the fits with a single power-law model leave similar tails. We also extract the pn background spectra from various locations of the CCD, and the tails persist as well. We further use a single power law model to fit all other available \xmm\ pn spectra of \pks, in particular the two observations performed in 2005 November 30 (orbit 1095) and 2007 April 22 (orbit 1349), only about half year before orbit 1171 and after orbit 1266. We find that these pn spectra are all typical of HBLs. As an example, Figure~\ref{fig:ratio}a also shows the pn spectrum and the data-to-model ratios for the observation in 2002 Nov 29-30 (orbit 0545, see Zhang et al. 2005, 2006 for the light curve) when the source was in an X-ray state similar to orbit 1171 and 1266. The residuals are completely distinguished from those of orbit 1171 and 1266-3, indicating that the X-ray spectra in 2006 indeed are not typical of HBLs. 

\begin{figure}
\centering
\includegraphics[scale=0.31, angle=270]{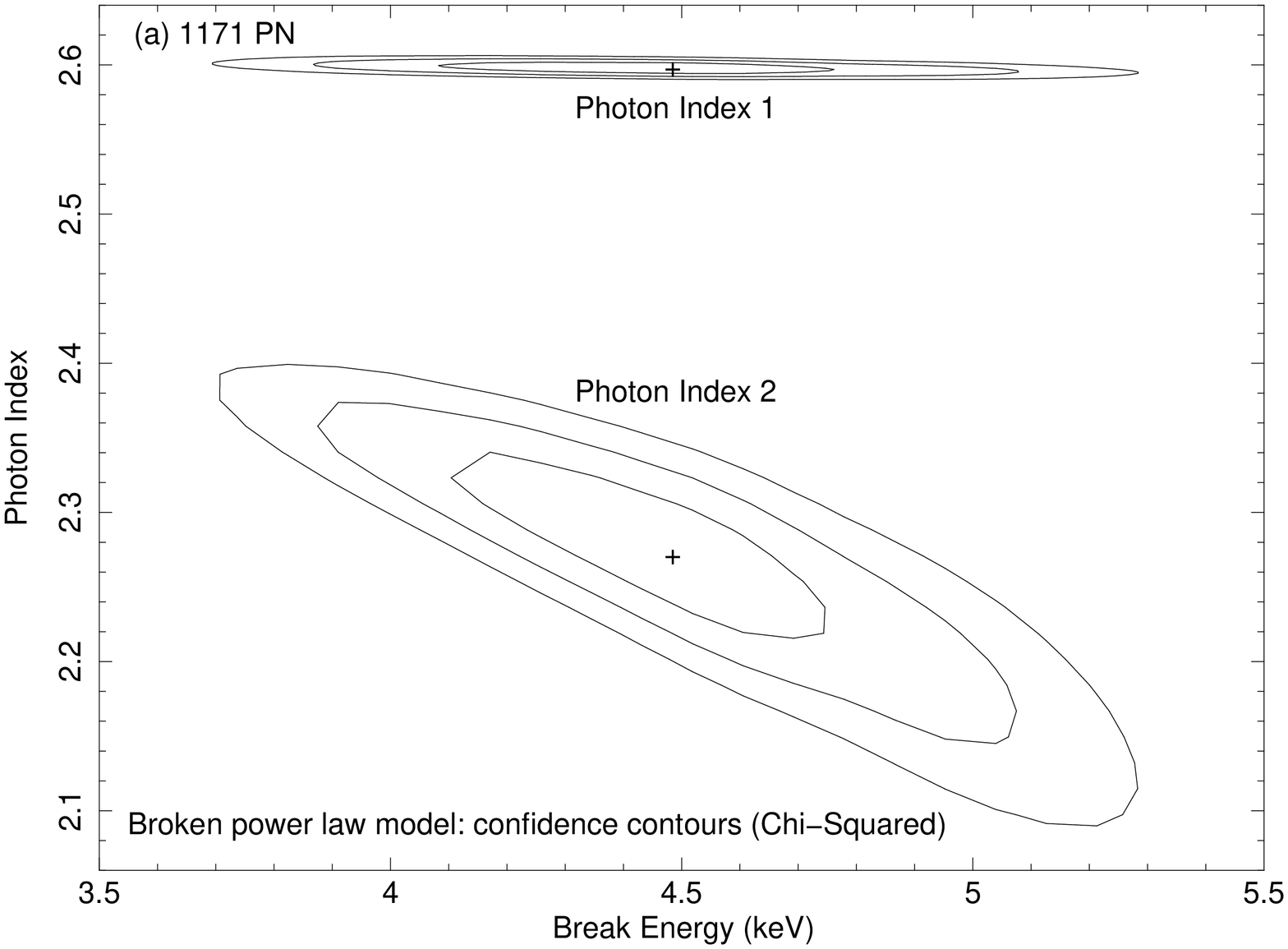}\includegraphics[scale=0.31, angle=270]{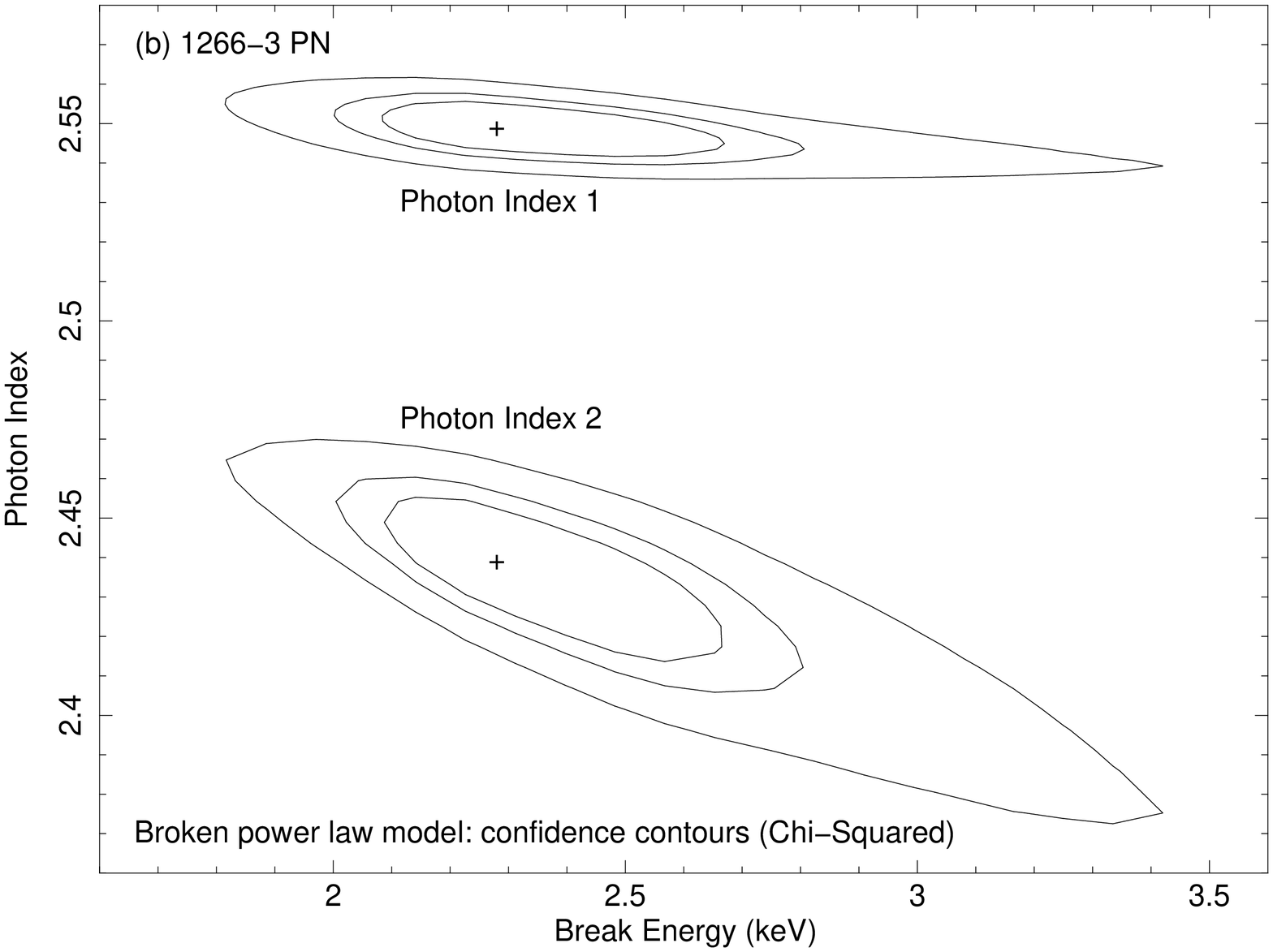}
\caption {\footnotesize  
Plot of the 68\%, 90\%, and 99\% confidence contours (from the inner to the outer curves, respectively) for the break energy and the soft (photon index 1) and the hard (photon index 2) X-ray photon index in the case of the broken power law best fitting model to the PN data of the orbit 1171 (a) and 1266-3 (b). These contours demonstrate that the soft and the hard X-ray photon index are completely disjoint at 99\% confidence level, confirming the hard tails of the concave X-ray spectra in 2006.  
} 
\label{fig:contour}
\end{figure}

Excellent spectral fits are achieved with a broken power-law model, showing that the spectra are flatter in the hard than in the soft X-rays for orbit 1171 and 1266. The spectral hardening ($\Delta\Gamma \sim$~0.1--0.3) occurs at a break energy of $\sim$~3--5~keV. This is sharply opposite to the spectral softening ($\Delta \Gamma \sim 0.23$) at a break energy of $\sim$~2~keV for orbit 0545. We present in Figure~\ref{fig:ratio}b the pn, mos1 and mos2 count spectra with the best broken power-law models and the data-to-model ratios for orbit 1171. Moreover, Figure~\ref{fig:contour} plots the 68\%, 90\%, and 99\% confidence level contours for the break energy and the soft/hard X-ray photon index under the best fits with the broken power law model to the PN spectra of the orbit 1171 and 1266-3. The contours demonstrate that the soft and the hard X-ray photon index are completely separate at 99\% confidence level, confirming the hard tails of the concave X-ray spectra of the source in 2006. The similar contours also show that the soft and the hard X-ray photon index are completely disjoint at 99\% confidence level for all other cases.

The logarithmic parabolic model, often used to describe continuously downward-curved X-ray spectra of HBLs (e.g., Massaro et al. 2004), also presents good fits. The curvature index is $0.22\pm0.01$ for the pn spectrum of orbit 0545, indicating downward-curved spectral shape. However, the curvature index is $-0.07\pm0.01$ and $-0.09\pm0.01$ for the pn spectra of orbit 1171 and 1266, respectively, suggesting that the spectra are weakly up-ward curved and in surpport with the concave spectral shape derived from the broken power-law fits.

                   %--------------------------------%

\section{Optical Data and Spectral Energy Distribution}\label{sec:sed}

The OM observations do not provide useful information for short-term variability of \pks\ in the optical-UV bands because only one or two images were obtained for each filter with exposure time of 1000--4400~s during each orbit. However, they do show long-term optical and UV variability over timescales of years. More importantly, the OM and X-ray observations present quasi-simultaneous optical-UV-X-ray SEDs that are precious to understand the long-term evolution of synchrotron spectra of the source. Figure~\ref{fig:sed} plots the SEDs for orbit 0545, 1171, and 1266-3, respectively. The X-ray SEDs are derived from the average pn spectra that are unabsorbed and unfolded with the best broken power-law model. The OM fluxes are dereddened by using the extinction formula of Cardelli et al. (1989). A 10\% error is assigned to each OM flux to account for the uncertainties of both the flux conversion and extinction correction. It is worth noting, during each orbit, that the OM observations at the six wavelengths are not simultaneous, and the PN and OM observations do not cover the same length of time (in fact, the OM exposure times are much shorter than the pn ones). As we are concentrated on the long term evolution of the SEDs of the source, the average pn fluxes are used for constructing the quasi-simultaneous SEDs. 

The X-ray SEDs of \pks\ clearly show an upturn in the high energy X-ray band for orbit 1171 and 1266-3. It is also obvious that the source goes through strong evolution of synchrotron emission spectra, manifested by very different amplitude of flux variability in different energy bands. The optical-UV fluxes increase by a factor of $\sim$~1.6--2.0, depending on the wavelengths, over a time interval of about four years from orbit 0545 to 1266-3. However, the 0.6--10~keV X-ray flux increases only $\sim$~38\% over the same time period even though the X-ray variability amplitude strongly increases with increasing energy. Similar flux variability amplitude of $\sim$~50\% occurs in the optical-UV and X-rays over a timescale of about half year from orbit 1171 to 1266-3. 

                    %---------------------------------%

\section{Discussion}\label{sec:disc}

The X-ray spectra of HBLs are characterized by convex shape or by continuously down-ward curved shape in some cases. This is the signature of the high energy tail of the synchrotron emission as the result of the energy-dependent particle acceleration and cooling (e.g., Massaro et al. 2004). Previous X-ray observations by various X-ray telescopes showed that the X-ray spectra of \pks\ are convex at the X-ray energy range below 10~keV (e.g., Kataoka et al. 2000; Zhang et al. 2002). Though numerous \xmm\ observations of the source exhibited the convex X-ray spectra as well (Foschini et al. 2006a; Massaro et al. 2008), the two \xmm\ observations in 2006 we analyzed here clearly show that the X-ray spectra of the source are concave. A broken power-law fit to the 0.6--10~keV spectra yield a break energy around 3--5~keV, with a spectral flattening of $\Delta\Gamma \sim$~0.1--0.3 from the soft to hard X-rays. However, such break energy usually occurs around energy of several tens of keV as revealed by the broad-band \sax\ X-ray spectra of \pks\ (Giommi et al. 1998; Chiappetti et al. 1999). This paper thus presents the first evidence for the concave X-ray spectra below 10~keV for a typical HBL.

In the X-rays below 10~keV, the concave X-ray spectra are a common feature of LBLs. This has been interpreted as a mixture of two spectral components. A steep and variable spectral component, attributed to the synchrotron emission from the high-energy tail of an electron distribution, dominates at the soft X-rays, while a flat but less variable spectral component, associated with the IC scattering of the low-energy side of the same electron distribution off the synchrotron (and/or external in some cases) photons, is dominant at the hard X-rays. This explanation is supported by the fact that the transiting energy between the two spectral component increases with increasing X-ray flux (e.g., S5~0716+714: Ferrero et al. 2006; Foschini et al. 2006b; BL Lacertae: Ravasio et al. 2003). Moreover, in very high states, the synchrotron emission from S5~0716+714 can be dominant up to 10~keV (Ferrero et al. 2006). 

In analogous to LBLs, we may attribute the concave X-ray spectra of \pks\ to a mixture of the two spectral component as well: the low energy X-ray emission, whose spectrum is steep ($\Gamma \sim 2.6$) , originates completely from the high enegy end of the synchrotron emission, but the high energy X-ray emission, whose spectrum is flatter than the soft one, should be a mixture of the most energetic end of the synchrotron emission and the low energy side of the IC emission. The IC emission whose spectrum is flat, gradually becomes important with increasing energy. However, the high energy X-ray spectra is steep ($\Gamma \sim 2.3-2.5$), implying that the high energy X-ray emission is still dominated by the synchrotron emission. The contamination of the IC emission flattens a bit the high energy X-ray spectra. If it is dominated by the IC component, the high energy X-ray spectra should be flat similar to LBLs (i.e., $\Gamma \sim 1.7$; e.g., Ferrero et al. 2006). Our analyses of all available \xmm\ observations show that \pks\ was in relatively low X-ray states in 2006, indicating that the emergence of the IC component may be realted to its low X-ray state. However, in other epochs with similar low X-ray states (e.g., orbit 0545 shown in this paper), the X-ray spectra of the source are still convex. Another possibility to account for the flatter but steep high energy X-ray spectra is that the X-ray emission of \pks\ in 2006 may consist of two synchrotron components with different spectral slopes. 

\begin{figure}
\centering
\includegraphics[scale=0.35, angle=270]{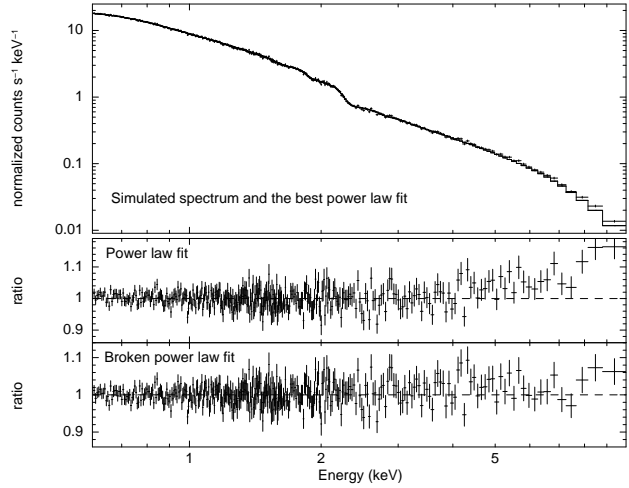}
\caption {\footnotesize  
Top panel shows the simulated X-ray spectrum and the best power law fit. The simulated spectrum is derived with a model consisting two power law components with photon index $\Gamma = 2.5$ and $\Gamma = 1.7$, respectively. The two components are assumed to contribute equally to the flux at 30~keV. The simulated spectrum is then convolved with orbit 1171 pn response and auxiliary files and the Galactic neutral hydrogen absorption, and the flux is also normalized to the orbit 1171 pn one with photon counting statistics of 60000~s exposure time. The best power law fit (middle panel) shows the hard tail of the simulated spectrum. The best broken power law fit (bottom panel) demonstrates the concave shape of the simulated spectrum, which is flatter in the hard ($\Gamma = 2.44\pm0.01$) than in the soft ($\Gamma = 2.31\pm0.05$) X-ray band with a break energy at $3.93\pm0.68$~keV. }
\label{fig:fake}
\end{figure}

With XSPEC 12.3.1, we simulate a spectrum consisting of two power law components: one with $\Gamma = 2.5$ represents the steep synchrotron component, and another one with $\Gamma = 1.7$ is for the flat IC component. We normalize the relative importance of the two components such that they contribute equally to the observed flux at 30~keV. The simulated spectrum is then folded with the pn response and auxiliary files of orbit 1171, and normalized to the pn 0.6--10~keV flux and photon counting statistics (with 60000~s exposure time) of orbit 1171 as well. We then group and fit the simulated spectrum in the exact same way as the observed spectra (\S~\ref{sec:spectra}), which is presented in Figure~\ref{fig:fake}. All the fits present very similar results and fit statistic as the observed ones. The single power law fit gives $\Gamma = 2.431\pm0.004$ ($\chi^2_{\nu}=1.08$, dof=1276 and probability=2.3\%) and leaves similar hard tail above $\sim 4$~keV (middle panel of Figure~\ref{fig:fake}). The broken power law fit results in $\Gamma_1 = 2.441\pm0.005$ and $\Gamma_2 = 2.309\pm0.05$ with break energy at $3.93\pm0.68$~keV ($\chi^2_{\nu}=1.04$, dof=1274 and probability=13.2\%), and the logarithmic parabolic model yields the curvature index of $-0.045\pm0.006$ and $\Gamma=2.449\pm0.004$ at 1~keV ($\chi^2_{\nu}=1.05$, dof=1275 and probability=9.0\%). The simulation thus demonstrates that the concave spectra with steep photon index at the hard X-rays is very likely a mixture of a steep synchrotron component dominating below 10~keV and a flat IC one domianting above 10~keV. However, the IC component is effectively present and perceived as a hardening of the synchrotron spectrum between $\sim 4$ and 10~keV. 

\begin{figure}
\epsscale{0.45}
\plotone{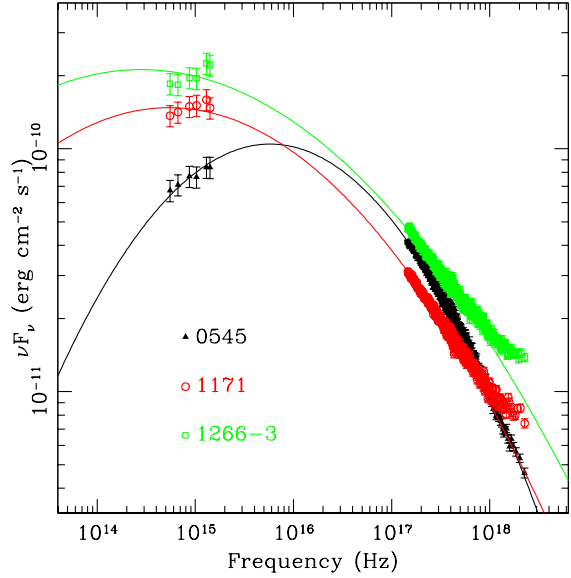}
\caption { \footnotesize 
The simultaneous optical-UV-X-ray SEDs at three epochs. The solid lines are the best parabolic fits to the data between the optical and 2~keV range, which is then extrapolated to 10~keV. For orbit 1171 and 1266-3, the deviation of the hard X-ray SED from the synchrotron SED (i.e., parabola) suggests possible contamination of the IC component. 
}
\label{fig:sed}
\end{figure}

The simultaneous optical-UV-X-ray SEDs show that \pks\ is more variable in the optical-UV bands than in the X-rays, implying strong evolution of the synchrotron spectra from orbit 0545 to 1171 and 1266-3. Although the synchrotron peak is not directly perceived due to a broad data gap between the optical-UV and X-rays, we may fit the SEDs with a parabola to phenomenologically determine the synchrotron peak frequency and flux. The fit is restricted to the optical-UV and 0.6--2~keV range to avoid putative contamination of the IC emission above $\sim$~3~keV. The parabola from the best fit is then extended to the whole observed X-ray range, which is plotted in Figure~\ref{fig:sed} along with the SEDs, to show the degree of the IC contribution to the synchrotron component (i.e. the parabola). The high energy X-ray SED is clearly above the extrapolation of the parabola for orbit 1171 and 1266-3, while for orbit 0545 it is very well in agreement with the extrapolation of the parabola. The derived peak frequency ($\nu_{\mathrm{peak}}$) of the parabola is $5.78 \times 10^{15}$, $5.41 \times 10^{14}$ and $2.81 \times 10^{14}$~Hz, and the corresponding peak flux, $(\nu F_{\nu})_{\mathrm{peak}}$, is $1.04 \times 10^{-10}$, $1.47 \times 10^{-10}$ and $2.12 \times 10^{-10} ~ \mathrm{erg ~ cm^{-2} ~ s^{-1}}$ for orbit 0545, 1171 and 1266-3, respectively. The peak flux is thus anti-correlated with the peak frequency: the lower the peak frequency, the higher the peak flux. The down-ward shift of the synchrotron peak to the optical wavelengths might be the reason that the IC component become visible below 10~keV for orbit 1171 and 1266-3. However, this behaviour is inconsistent with the positive correlation between the frequency and flux of the synchrotron peak found in some X-ray bright TeV HBLs (Tramacere et al. 2007; Massaro et al. 2008). In fact, Massaro et al. (2008) already noticed that \pks\ behaves differently. This new discovery warrants further more simultaneous multiwavelength observations with the EUV range involved in particular and theoretical considerations. 

                 %----------------------%

\acknowledgments

I thank the anonymous referee for the constructive suggestions and comments that improve significantly the paper. This research is based on observations obtained with \xmm, an ESA science mission with instruments and contributions directly funded by ESA Member States and NASA. This work is conducted under the Key Project of Chinese Ministry of Education (NO 106009), Project 10473006 and 10733010 supported by NSFC, Project sponsored by the Scientific Research Foundation for the Returned Overseas Chinese Scholars, State Education Ministry, and Project for top scholars sponsored by Tsinghua University.

		  %-----------------------%  

                      %----------------------------%

\begin{deluxetable}{llrccccccc}
\tabletypesize{\scriptsize}
\tablecolumns{9}
\tabcolsep 3pt
\tablewidth{0pc}
\tablecaption{Observational Journal of \pks\ with \xmm\ EPIC PN and MOS}
\tablehead{
\colhead{Obs. ID} &\colhead{Orbit} &\colhead{Date (UT)} &\colhead{Detector} &\colhead{Mode\tablenotemark{a}} &\colhead{Filter} &\colhead{Duration (s)} &\colhead{Live Time (s)} &\colhead{Count Rate\tablenotemark{b}} 
}
\startdata

0158961401 &1171   &2006 May 1 12:31:58-06:24:50\tablenotemark{c}  &PN   &SW &Medium &64372  &45108 &12.2 \\
           &1171   &           12:26:48-06:22:50\tablenotemark{c}  &MOS1 &SW &Medium &64563  &62638 &3.24 \\
           &1171   &           12:26:48-06:22:54\tablenotemark{c}  &MOS2 &SW &Medium &64568  &62637 &3.23 \\

0411780101 &1266-1 &2006 Nov 7 00:31:25-08:50:03  &PN   &SW &Thin   &29918  &20869 &17.3 \\
           &1266-2 &           08:52:19-18:35:53  &PN   &SW &Medium &35015  &24356 &18.8 \\
           &1266-3 &           18:38:36-04:24:57\tablenotemark{c}  &PN   &SW &Thick  &35179  &24626 &16.6 \\
           &1266   &           00:23:33-04:22:06\tablenotemark{c}  &MOS1 &SW &Medium &100714 &97288 &2.66 \\
           &1266   &           00:23:34-04:22:11\tablenotemark{c}  &MOS2 &SW &Medium &100718 &97285 &2.65 \\
0124930601 &0545   &2002 Nov 29 23:32:52-15:20:17\tablenotemark{c} &PN &SW &Thick &56845 &39794 & 13.0 \\      
\enddata    
\tablenotetext{a}{SW indicates imaging small window.}
\tablenotetext{b}{Background subtracted mean count rate in the 0.6--10~keV energy band.}
\tablenotetext{c}{Next day.}
\label{tab:epic}
\end{deluxetable}

\begin{deluxetable}{lcccccc}
\tabletypesize{\scriptsize}
\tablecolumns{7}
\tabcolsep 3pt
\tablewidth{0pc}
\tablecaption{Observational Results of \pks\ with \xmm\ OM (Imaging Mode)}
\tablehead{
\colhead{Filter} &\colhead{Wavelength (nm)} &\colhead{Data (UT)} &\colhead{Exp. Time (s)} &\colhead{Count Rate\tablenotemark{a}} &\colhead{Magnitude\tablenotemark{b}} &\colhead{Flux\tablenotemark{c}}
}
\startdata
\multicolumn{7}{c}{Orbit 1171 (2006-05-01)} \\
V    &543 &12:34:59-12:51:40 &1000 &$95.08\pm0.58$ &$13.02\pm0.01$ &$2.38\pm0.02$ \\
U    &344 &12:56:48-13:13:26 &1000 &$202.37\pm1.14$ &$12.49\pm0.01$ &$3.93\pm0.02$ \\
B    &450 &13:18:32-13:35:13 &1000 &$224.98\pm1.48$ &$13.39\pm0.01$ &$2.81\pm0.02$ \\ 
UVW1\tablenotemark{d} &291 &17:37:40-18:49:49 &4000 &$96.63\pm0.32$ &$12.24\pm0.01$ &$4.66\pm0.02$ \\
UVM2 &231 &19:37:55-20:44:35 &4000 &$26.24\pm0.08$ &$12.23\pm0.01$ &$5.80\pm0.02$ \\
UVW2 &212 &20:49:42-21:56:23 &4000 &$10.41\pm0.05$ &$12.32\pm0.01$ &$5.94\pm0.03$ \\
\multicolumn{7}{c}{Orbit 1266 (2006-11-07)} \\
V    &543 &08:55:01-09:11:41 &1000 &$128.76\pm0.77$ &$12.69\pm0.01$ &$3.22\pm0.02$ \\
U    &344 &09:46:49-10:03:29 &1000 &$265.76\pm1.39$ &$12.20\pm0.01$ &$5.16\pm0.03$ \\
B    &450 &10:08:36-10:25:17 &1000 &$291.31\pm2.47$ &$13.11\pm0.01$ &$3.64\pm0.03$ \\
UVW1\tablenotemark{d} &291 &14:27:42-16:46:31 &8000 &$124.82\pm0.29$ &$11.96\pm0.01$ &$6.02\pm0.01$ \\
UVM2 &231 &17:34:37-18:41:18 &4000 &$37.05\pm0.09$ &$11.85\pm0.01$ &$8.19\pm0.02$ \\
UVW2 &212 &18:46:24-19:53:04 &4000 &$15.21\pm0.06$ &$11.91\pm0.01$ &$8.67\pm0.03$ \\
\multicolumn{7}{c}{Orbit 0545 (2002-11-30)} \\
V    &543 &00:15:02-00:31:41 &1000 &$46.62\pm0.20$ &$13.79\pm0.01$ &$1.16\pm0.01$ \\
U    &344 &00:36:49-00:56:49 &1200 &$104.38\pm0.57$ &$13.21\pm0.01$ &$2.03\pm0.01$ \\
B    &450 &01:01:55-01:18:35 &1000 &$112.74\pm0.72$ &$14.14\pm0.01$ &$1.45\pm0.01$ \\
UVW1 &291 &03:54:25-04:44:25 &3000 &$47.86\pm0.10$ &$13.00\pm0.01$ &$2.28\pm0.01$ \\
UVM2 &231 &04:49:32-05:39:32 &3000 &$14.26\pm0.07$ &$12.89\pm0.01$ &$3.14\pm0.02$ \\
UVW2 &212 &05:44:38-06:57:58 &4400 &$5.83\pm0.04$ &$12.95\pm0.01$ &$3.33\pm0.02$ \\ 
\enddata  
\tablenotetext{a}{Backgound subtracted count rate.}
\tablenotetext{b}{Instrumental magnitude.} 
\tablenotetext{c}{Flux in unit of $10^{-14}~{\rm ergs ~ cm^{-2} ~ s^{-1}~\AA^{-1} }$ is converted from the count rate using the conversion factor of white dwarf standard star.}
\tablenotetext{d}{Two images are combined.}
\label{tab:om}
\end{deluxetable}

\begin{deluxetable}{lcccccccc}
\tablecolumns{9}
\tabletypesize{\scriptsize}
\tabcolsep 5pt
\tablewidth{0pt}
\tablecaption{Results of the X-ray Spectral Fits of \pks\tablenotemark{a}}
\tablehead{
\colhead{Orbit} &\colhead{$\Gamma_1$\tablenotemark{b}} &\colhead{$E_{\rm break}$/$b$\tablenotemark{c}} &\colhead{$\Gamma_2$} &\colhead{K\tablenotemark{d}} &\colhead{$\chi^2_{\nu}$/dof} &\colhead{Prob.} &\colhead{F$_{2-10~{\rm keV}}$\tablenotemark{e}} &\colhead{F$_{0.6-10~{\rm keV}}$\tablenotemark{f}}
}
\startdata
\multicolumn{9}{c}{Power law (PL) model}\\
1171/PN   &$2.59\pm0.01$ &-- &-- &$1.44\pm0.01$ &1.11/1272 &0.003 &1.60 &4.28\\

1171/MOS1 &$2.55\pm0.01$ &-- &-- &$1.47\pm0.01$ &1.12/375  &0.053 &1.72 &4.46\\

1171/MOS2 &$2.58\pm0.01$ &-- &-- &$1.53\pm0.01$ &1.12/382  &0.053 &1.72 &4.58\\

1266-1/PN &$2.53\pm0.01$ &-- &-- &$1.91\pm0.01$ &1.03/1127 &0.241 &2.28 &5.84\\

1266-2/PN &$2.52\pm0.01$ &-- &-- &$2.11\pm0.01$ &1.01/1226 &0.412 &2.57 &6.51\\

1266-3/PN &$2.52\pm0.01$ &-- &-- &$2.24\pm0.01$ &1.12/1245 &0.003 &2.73 &6.91\\

1266/MOS1 &$2.48\pm0.01$ &-- &-- &$2.09\pm0.01$ &1.22/406  &0.001 &2.70 &6.47\\

1266/MOS2 &$2.51\pm0.01$ &-- &-- &$2.21\pm0.01$ &1.18/402  &0.008 &2.74 &6.87\\

0545/PN   &$2.75\pm0.01$ &-- &-- &$2.21\pm0.01$ &1.72/1134 &0.000 &1.65 &5.13\\

\cline{1-9}
\multicolumn{9}{c}{Broken power law (BKPL) model}\\

1171/PN  &$2.60\pm0.01$ &$4.49^{+0.37}_{-0.47}$ &$2.27^{+0.08}_{-0.07}$ &$1.44\pm0.01$ &1.00/1270 &0.511 &1.65 &4.33 \\

1171/MOS1 &$2.56\pm0.01$ &$3.04^{+0.65}_{-0.38}$ &$2.46^{+0.04}_{-0.06}$ &$1.47\pm0.01$ &1.09/373 &0.112 &1.76 &4.50 \\

1171/MOS2 &$2.59\pm0.01$ &$4.35^{+0.44}_{-0.54}$ &$2.26^{+0.12}_{-0.13}$ &$1.53\pm0.01$ &1.04/380 &0.307 &1.79 &4.65 \\

1266-1/PN &$2.54\pm0.01$ &$3.31^{+0.66}_{-0.72}$ &$2.45^{+0.03}_{-0.04}$ &$1.90\pm0.01$ &1.01/1125 &0.433 &2.31 &5.87 \\

1266-2/PN &$2.53\pm0.01$ &$4.46^{+0.49}_{-0.71}$ &$2.33^{+0.07}_{-0.07}$ &$2.11\pm0.01$ &0.96/1224 &0.818 &2.61 &6.56  \\

1266-3/PN &$2.55\pm0.01$ &$2.29^{+0.39}_{-0.21}$ &$2.44^{+0.02}_{-0.03}$ &$2.24\pm0.01$ &1.03/1243 &0.243 &2.79 &6.97 \\

1266/MOS1 &$2.49\pm0.01$ &$3.20^{+0.44}_{-0.38}$ &$2.37^{+0.04}_{-0.05}$ &$2.09\pm0.01$ &1.17/404 &0.011 &2.77 &6.69  \\ 

1266/MOS2 &$2.52\pm0.01$ &$3.07^{+0.30}_{-0.30}$ &$2.41^{+0.04}_{-0.04}$ &$2.21\pm0.01$ &1.12/400 &0.044 &2.80 &6.93 \\ 

0545/PN   &$2.67\pm0.01$ &$1.70^{+0.08}_{-0.07}$ &$2.90\pm0.01$ &$1.88\pm0.01$ &1.02/1132 &0.289 &1.56 &5.06 \\

\cline{1-9}
\multicolumn{9}{c}{Logarithmic parabolic (LP) model}\\

1171/PN &$2.61\pm0.01$ &$-0.07\pm0.01$ &-- &$1.43\pm0.01$ &1.05/1271 &0.115 &1.64 &4.32 \\

1171/MOS1 &$2.55\pm0.01$ &$-0.02\pm0.03$ &-- &$1.47\pm0.01$ &1.12/374 &0.055 &1.73 &4.48 \\

1171/MOS2 &$2.59\pm0.01$ &$-0.02\pm0.03$ &-- &$1.53\pm0.01$ &1.12/381 &0.055 &1.73 &4.60 \\

1266-1/PN &$2.55\pm0.01$ &$-0.05\pm0.01$ &-- &$1.90\pm0.01$ &1.01/1126 &0.447 &2.32 &5.87 \\

1266-2/PN &$2.55\pm0.01$ &$-0.06\pm0.01$ &-- &$2.10\pm0.01$ &0.96/1225 &0.820 &2.62 &6.56 \\

1266-3/PN &$2.56\pm0.01$ &$-0.09\pm0.01$ &-- &$2.23\pm0.01$ &1.02/1244 &0.332 &2.80 &6.98 \\

1266/MOS1 &$2.49\pm0.01$ &$-0.03\pm0.02$ &-- &$2.09\pm0.01$ &1.22/405 &0.002 &2.73 &6.65 \\

1266/MOS2 &$2.52\pm0.01$ &$-0.02\pm0.02$ &-- &$2.21\pm0.01$ &1.18/401 &0.009 &2.75 &6.89 \\

0545/PN   &$2.67\pm0.01$ &$0.22\pm0.01$  &-- &$1.90\pm0.01$ &1.00/1133 &0.513 &1.56 &5.05 

\enddata
\tablenotetext{a}{The fits are performed in the 0.6--10~keV energy band. The neutral hydrogen absorption column density is fixed to the Galactic value ($N_{\rm H} = 1.36\times 10^{20}~ {\rm cm^{-2}}$). The quoted errors are $90\%$ confidence level for one interesting parameter.}
\tablenotetext{b}{$\Gamma_1$ is the photon index at 1~keV in the LP model. The errors are smaller than 0.01.}
\tablenotetext{c}{$E_{\rm break}$ is the break energy in the BKPL model. $b$ is the curvature parameter in the LP model ($b>0$ means downward curved shape and $b<0$ upward cuved shape).}
\tablenotetext{d}{$K$ is the normalization factor in unit of $\rm{10^{-2}~ keV^{-1} ~ cm^{-2}~ s^{-1}}$. The errors are smaller than 0.01.}
\tablenotetext{e}{Unabsorbed 2--10~keV flux in unit of $\rm{10^{-11}~ergs ~ cm^{-2}~ s^{-1}}$.}
\tablenotetext{f}{Unabsorbed 0.6-10~keV flux in unit of $\rm{10^{-11}~ergs ~ cm^{-2}~ s^{-1}}$.}
\label{tab:fit}
\end{deluxetable}

                %------------------------------%

\end{document}